\documentstyle[12pt,aasms4]{article}

\def\deg{$^{\circ}\,$}

\begin{document}

\title{The Symmetry, Color, and Morphology of
                   Galaxies}

\author{CHRISTOPHER J. CONSELICE$^1$}
\vspace{5mm}

\affil{\centerline{Department of Astronomy and Astrophysics} \centerline{University of Chicago} \centerline{5640 S. Ellis Avenue} \centerline {Chicago, IL 60637} \centerline{chris@astro.wisc.edu}}

\altaffiltext{1}{Present Address: Department of Astronomy, University of Wisconsin, Madison, 475 N. Charter St., Madison, WI., 54706 }

\begin{abstract}
        The structural symmetry of forty-three face-on galaxy images in the R(650 nm) and J(450 nm) bands are measured to determine the usefulness of symmetry as a morphological parameter. Each galaxy image is rotated by $180$\deg and subtracted from the original to obtain a quantitative value for its structural symmetry.  The symmetry numbers computed for the sample are then compared with RC3 morphological types, color \& absolute blue magnitudes.   A strong correlation between color and symmetry is found, and the RC3 Hubble sequence is found to be one of increasing asymmetry.  The use of symmetry as a morphological parameter, and the possible causes of the asymmetries are discussed.

\end{abstract}

\clearpage

\section{INTRODUCTION}
     Since the beginning of extragalactic astronomy it has generally  been assumed that face-on spiral galaxies are for the most part axi-symmetric objects.  Recent studies using Fourier techniques have shown that deviations from symmetry exist for spiral galaxies (Rix \& Zaritsky 1995).  Evidence has also been accumulated that shows spirals as axi-asymmetric in hydrogen gas (Richter \& Sancisi 1994; Baldwin et al 1980).  Baldwin et al. (1980) concluded that the hydrogen gas asymmetries in their sample galaxies are long lived.  The more recent investigation by Richter \& Sancisi (1994) reached the same conclusion from  a much bigger sample.  Despite these efforts to investigate extragalactic asymmetries, attempts to quantify this measure in visual wavelengths for field galaxies are almost non-existent.  There have been, however, some attempts to  use asymmetry for morphological  purposes.

  Using the symmetry of spiral arms as a morphological indicator was begun by Elmegreen \& Elmegreen (1982 \& 1987).  In their approach, the apparent symmetry of the arms of flocculent and grand-design galaxies are classified according to the symmetry and continuity of  spirals arms.  In a separate study using the symmetric images of galaxies generated by rotation and subtraction, the underlying spiral structure is examined closely (Elmegreen, Elmegreen, \& Montenegro 1992).  Using this method it is shown that many spirals have hidden structures including  triple arm patterns.    

   As the classification of galaxies has been extended to include galaxies at  high redshift, such as those seen in deep Hubble Space Telescope images (Driver, et al. 1995; Odewahn, et al. 1996; Abraham, et al. 1996a), it is becoming apparent that normal classification procedures are inadequate for the faint irregular galaxies seen in these images.  One way to solve this problem is to  use automated classification through artificial neural networks (Odewahn, et al. 1996) to assign galaxies  classical morphological types.  Another possible method uses structural symmetry measurements for classification purposes.    Several studies have used this  approach to classify galaxies seen in the Hubble Deep Field  (Abraham et al. 1996b; van den Bergh et al. 1996).  Using asymmetry and the central concentration of the galaxian light these studies show morphological differences between the  Hubble Deep Field galaxies.  As more images of faint and distant galaxies are becoming accessible, the use of symmetry is becoming a powerful method of obtaining morphological information (Abraham et al. 1996b; Schade et al. 1995). 

   Before applying symmetry as a morphological indicator it is important to test it's usefulness and limitations on nearby galaxies.  In this paper it is shown that although limited,  symmetry is a powerful morphological indicator that can be used to estimate the classical morphological type, color and large scale physical features of galaxies.

\section{OBSERVATIONS}

     The sample of galaxies selected for this analysis were obtained from the Frei et al. (1996) public FTP site of galaxy images.   The galaxies in this catalog are large nearby high surface brightness galaxies of various morphologies and inclinations (Frei et al. 1996).  Out of this set of images, the ones in the R($650$nm) and J($450$nm) bands were used.  Of these only the ellipticals and face on spirals went through the analysis procedure.   The galaxies fitting these requirements  are listed in Table 1.  

     The images were taken with the 1.1m Hall telescope at Lowell observatory between March 24th and April 4th, 1989.  Each galaxy has an image in the R (650nm) and J(450nm) wavebands of the Gullixson et al. (1995) photometric system.  The CCD camera used was a 320 x 512 pixel RCA.  The time length of the exposures are 30 minutes for the R band and 45 minutes for the J band.  

    The sample consists of eight ellipticals or S0 galaxies, the remaining thirty five galaxies are face on spirals.  The spirals are evenly divided between early \& intermediate-types (Sa, Sb) and late-types (Sc, Sd).

\section{METHOD OF OBTAINING ASYMMETRY NUMBER}

\subsection{Method}

     The method for calculating the symmetry number A is simple, but must be carefully done.    Each image is first subtracted of background stars (Frei et al.  1996) and the sky background.  The center of the galaxy is then found by using the IRAF imcntr command, which roughly finds the brightest point in the center of the galaxy.  The galaxy image is next rotated by $180$\deg at it's center.  The rotated image is then subtracted from the original image with the two centers lined up. A residual map of all asymmetric components of the galaxy is then created.  This process is carried out separately for each image at both wavelengths.

    This residual map will have on average half its values positive and half negative.  The pixels in this residual map are then squared to obtain a residual squared map with all quantities positive.  An area which contains the galaxy is then summed of its pixel values in the squared image.  This sum is then divided by twice the sum of the squared pixels  of the corresponding region of the original image.  In formula: $$A^2\equiv\frac{\Sigma\frac{1}{2}(I_o-I_{180})^2} {\Sigma I_o^2}$$

 This ratio, which is never more than 1 and never less than 0 gives the degree of asymmetry of the galaxy.  A value of A=0 would correspond to a galaxy that is completely symmetric.  A value of A=1 would correspond to a galaxy which is completely asymmetric.   A completely symmetric galaxy is one where every point of light has a corresponding point of exactly the same brightness at the same distance from the center and $180$\deg away.  A completely asymmetric galaxy is one where every point of light has no corresponding brightness at the same distance from the center and $180$\deg away.  This asymmetry index  used in Abraham, et al. (1996b) is computed in a similar manner, but uses the sum of the absolute values of the subtracted pixels.  In this paper, we use the square root of the sum of the squares of the pixels.   The resulting values of A(J) and A(R) are shown in columns 4 and 5 of Table 1.

\subsection{Reliability and Limitations of the Method}

      There are a number of factors which could bias the results of a symmetry measurement.   If a galaxy is very distant then the resolution of that galaxy will be less than the resolution of a nearby galaxy using the same telescope.  The galaxy that is further away will appear smoother and the symmetry measurement will give a lower symmetry number than what would have been found if it were at a closer distance.  

   Likewise, when a galaxy is very close, it will appear more asymmetric due to the high resolution of the structure.    To test if these effects were present in the data, symmetry and distance are plotted in Figure 1.   The distances were taken from the ``Nearby Galaxy Catalog'' (Tully 1988).  A slight distance effect can be seen, which suggests some caution when applying this method to galaxies groups in different redshift ranges.   If the Virgo cluster galaxies at 17 Mpc are removed from the sample, the distance effect becomes more pronounced, hinting that symmetry may also be useful as a distance indicator.

\section{RESULTS}

    Based on the results in Table 1,  the J band images are found to be more asymmetric on average then those in the R band (Figure 2).   The breakdown in the symmetries are as follows: in the J band no galaxy has an A$>$0.25, with 45\% having an A$<$0.1.  In the  R band, no image has an A$>$0.17, with most, 54\% having A$<$0.1.   Most galaxies in our sample are therefore not extremely asymmetric.

  To see if symmetry correlates with other physical parameters, we plot the asymmetry numbers in our sample with morphological types and the integrated colors of the galaxies.  In Figure 3 the asymmetry numbers are plotted against morphological types.  Low numbers (-1,0 1,2) are early type systems, (E, S0, Sa, Sab) and higher numbers (3,4,5,6,7) are late type systems, (Sb, Sbc, Sc, Scd, Sd).  A trend is seen between the Hubble morphology and the symmetry.   As the galaxies in the sample get to later type spirals, the asymmetry parameter increases.  This suggests that the RC3 Hubble sequence is one of increasing optical asymmetry.   To determine if symmetry correlates with the absolute B magnitude (M) of our sample, we plot both physical parameters in Figure 4.   No correlation is found between these two features.  There is also no correlation between symmetry and M within each Hubble type.
 
    The symmetry-color plot (Figure 5) has a very strong correlation,  suggesting a fundamental relationship between these two physical parameters.  The degree of asymmetry of a galaxy can therefore be used as a measure of its global stellar populations.  Galaxies which are asymmetric have stellar populations that are blue, indicating that these galaxies consist of recently formed massive young stars.   The symmetric galaxies have a redder (B-V) color which suggests that the stellar population in these galaxies is relatively old.   The high asymmetries of these galaxies is due to patchy star formation through out the disk, and not due to a large global asymmetry due to misalignment of spiral arms.  Furthermore, the galaxies are almost always more symmetric in the redder R band (Figure 2), indicating that the  older stellar populations are smoothed out through time.  Any global asymmetries would also affect both the R and the J bands asymmetries equally.  Since the asymmetries are caused by recently formed stars, symmetry can also be used as a measure of the star formation rate (Conselice 1996).   Symmetry also weakly correlations with UV and H-$\alpha$ fluxes.

    The strong correlation between asymmetry and color can be used to find one when the other is not known.  Images of galaxies can be used to determine global physical features with the quantitative measurement of symmetry, and to infer color.   Likewise, an idea of a galaxy's morphology can be obtained by measuring its color if its image is not easily seen or resolved. 

\section{SUMMARY AND DISCUSSION}

    As the use of symmetry to obtain morphological and physical information on galaxies at high redshifts is bound to increase, it is desirable to obtain relations between it and other physical parameters in nearby field galaxies.  A  strong correlation between the structural symmetry and color (Figure 5) suggests a fundamental relationship between these two physical parameters.  The correlation between symmetry and RC3 Hubble morphological type (Figure 3) suggest that the RC3  Hubble morphology is a sequence of increasing asymmetry.   

   Using the symmetry methods presented in this paper, physical parameters of galaxies which would otherwise be impossible to obtain could be reasonably estimated.   Compact and distant galaxies whose disks are not easily resolved can  have their large scale physical features and their global morphology deciphered.  Symmetry is therefore a limited but powerful method of deriving physical and morphological information that would otherwise be unattainable.

   It is a pleasure to thank Richard G. Kron for invaluable assistance and advice throughout the completion of this study.  Jay Gallagher and  Debbie Elmegreen looked over  early versions of the manuscript and made several valuable suggestions.   Marianna Takamiya,  Rhodri Evans, and Al Harper also contributed helpful discussions and advice.   I would also like to thank the Frei et al. (1996) group for making their galaxy images publicly available.  Support from the University of Wisconsin, Madison during the preparation of this paper is also graciously acknowledged.



\vspace{3cm}

FIGURE CAPTIONS:

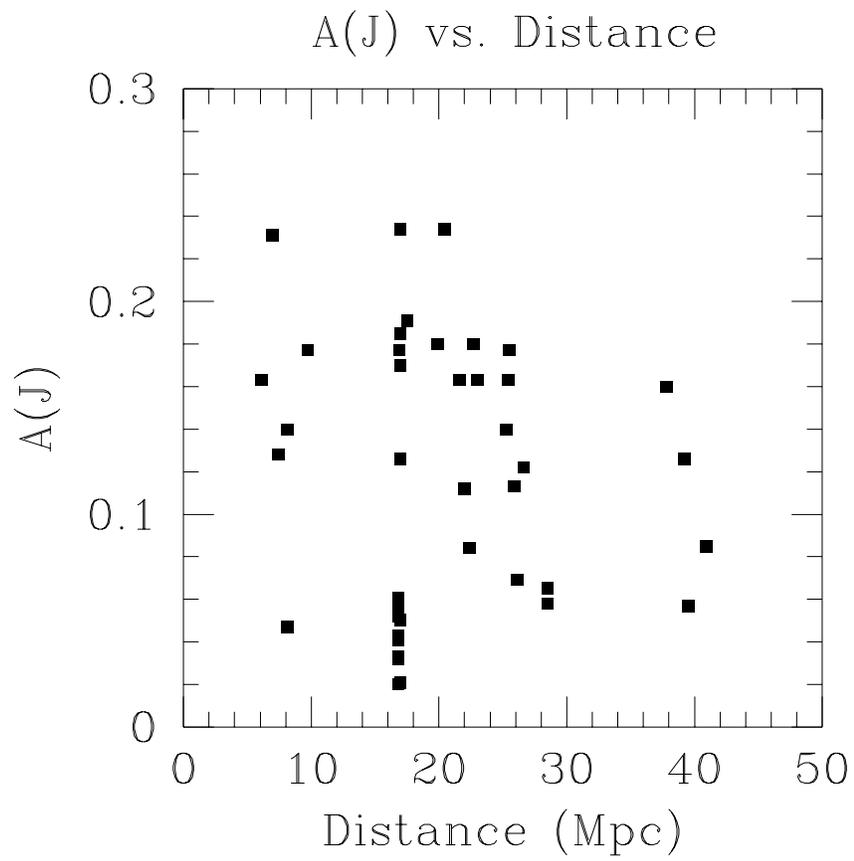
Fig.1 --- The symmetry-distance relation shows that for our sample distance effects are not heavily biasing our sample.  A slight decrease might indicate that only low redshift galaxies could be included in this sample to ensure no smoothing bias caused by distant galaxies.

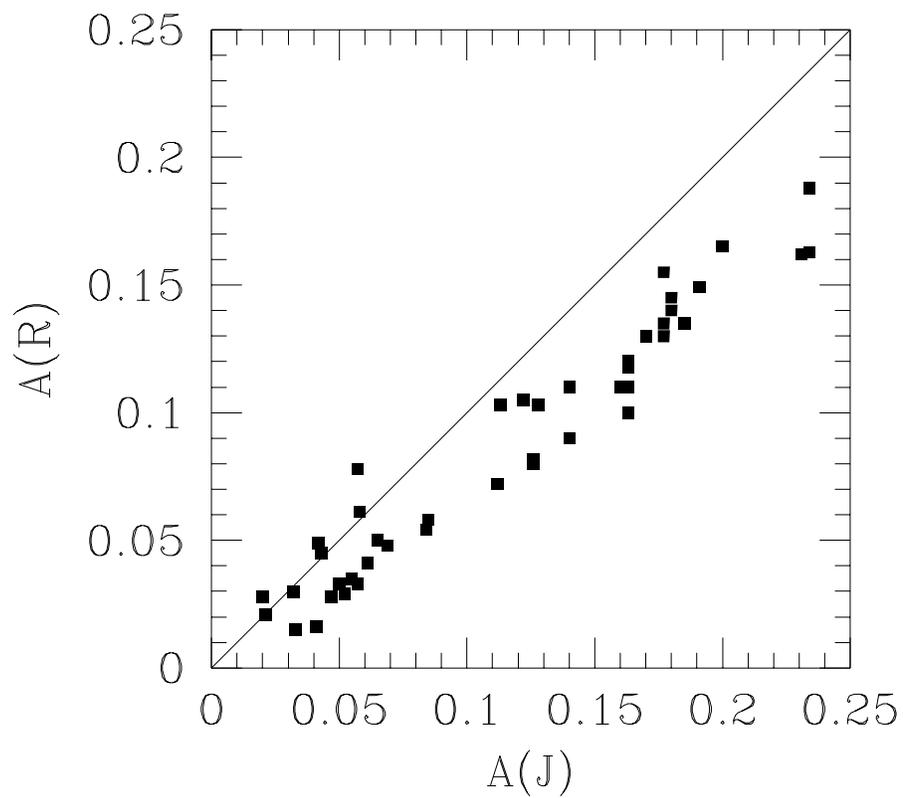
Fig.2 --- The relation between the symmetry numbers in each of the two wave bands, R and J used in the analysis.   Symmetry numbers computed in  the J band images are almost always more asymmetric than the R band images.  This show the old populations in a galaxy are  more evenly distributed than the younger stars.

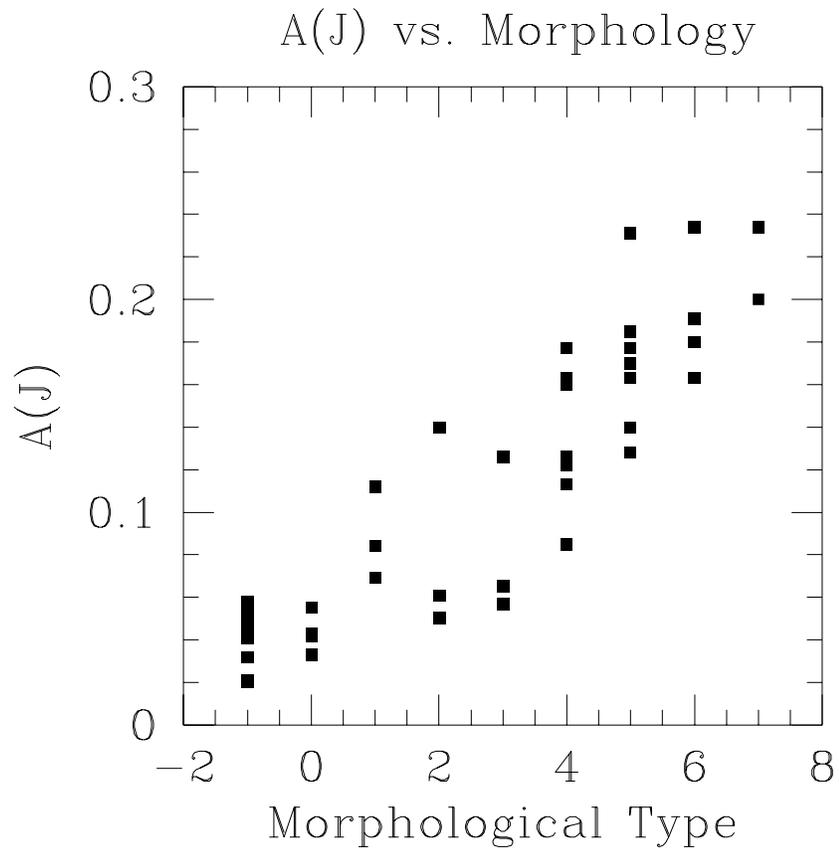
Fig.3 ---  The relationship between the morphology as given in the RC3 catalog
(de Vaucouleurs et al, 1991) and the symmetry numbers in the J band.  The numbers represent the various morphological types (-1,0,1,2,3,4,5,6,7 are E, S0, Sa, Sab, Sb, Sbc, Sc, Scd, Sd, respectively).  A trend is seen between the morphological type and the symmetry number, suggesting that the RC3 Hubble sequence is one of increasing asymmetry. A similar relationship is found for the R band.

Fig.4 --- The absolute blue magnitudes of our galaxy sample plotted with the asymmetry numbers.  No correlation is found between these two parameters.

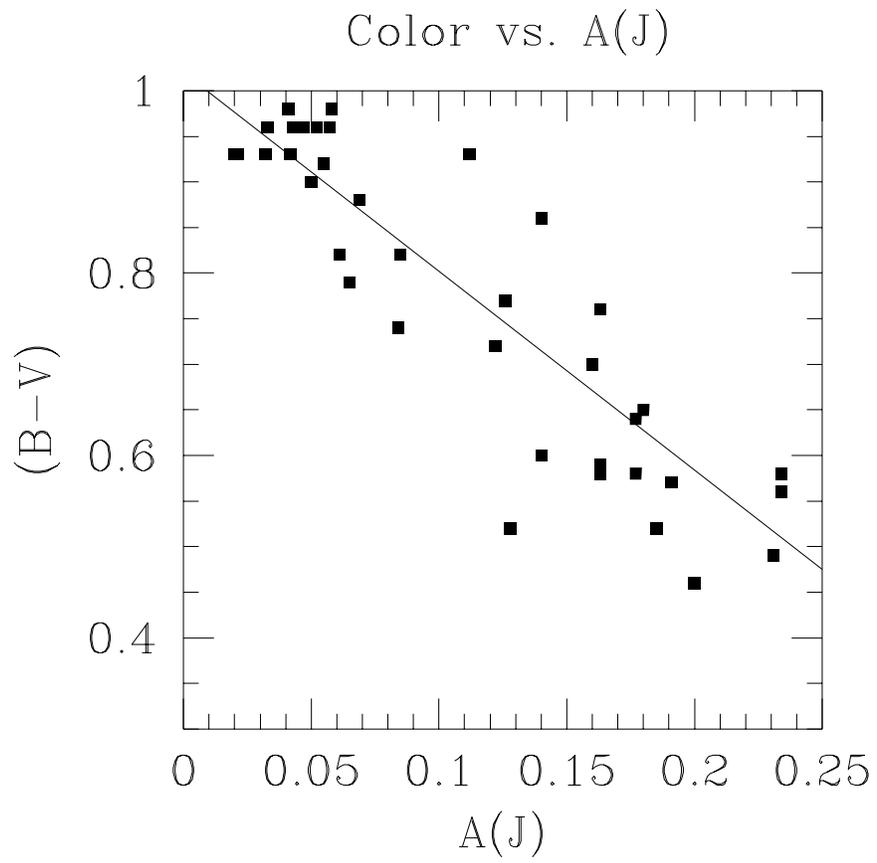
Fig.5 --- The symmetry-color relation.  An excellent correlations is found
between these two physical parameters suggesting that this relationship is
a fundamental one to galaxies.

\clearpage

\begin{table}
\begin{center}

Table 1: Sample\\
\vspace{1ex}
\begin{tabular} {rlllllll}
\hline
 & Galaxy & Type &  A(J)  &  A(R) & (B-V) & D(Mpc) & M$_B$ \\
\hline
\hline
1.& NGC 4621 & E5 & 0.020 & 0.028 & 0.93 & 16.8 & -20.45\\
2.& NGC 4636 & E0 - 1 & $0.021$ & $0.021$ & $0.93$ & 17.0 & -20.68 \\
3.& NGC 4406 & --- & 0.032 & 0.03 & 0.93 & 16.8 & -21.15\\
4.& NGC 4472 & E2/S0(2) & 0.033 & 0.015 & 0.96 & 16.8 & -21.85\\
5.& NGC 4374 & E1 & 0.041 & 0.016 & 0.98 & 16.8 & -20.95\\
6.& NGC 4340 & SB0(r) & 0.042 & 0.0049 & 0.93 & 16.8 & -19.17\\
7.& NGC 4477 & SB(s)0 & 0.043 & 0.0045 & 0.96 & 16.8 & -19.87 \\
8.& NGC 3379 & E1 & 0.05 & 0.033 & 0.96 & 8.1 & -19.39\\
9.& NGC 2775 & SA(r)ab & 0.052 & 0.029 & 0.9 & 17.0 & -20.13\\
10.& NGC 4365 & E3 & 0.055 & 0.035 & 0.96 & 16.8 & -20.52\\
11.& NGC 4754 & SB(r)0- & 0.057 & 0.078 & 0.92 & 16.8 & -19.74\\
12.& NGC 4593 & SB(rs)b & 0.057 & 0.033 & --- & 39.5 & -21.56\\
13.& NGC 4486 & E2 & 0.058 & 0.061 & 0.9 & 16.8 & -21.68\\
14.& NGC 5813 & E1 - 2 & 0.061 & 0.041 & 0.98 & 28.5 & -20.77\\
15.& NGC 4450 & SA(s)ab & 0.065 & 0.05 & 0.82 & 16.8 & -20.36\\
16.& NGC 5850 & SB(r)b & 0.069 & 0.048 & 0.79 & 28.5 & -20.69\\
17.& NGC 5701 & SB(rs)0/a & 0.084 & 0.054 & 0.88 & 26.1 & -20.35\\
18.& NGC 2985 & SA(rs)a & 0.085 & 0.058 & 0.74 & 22.4 & -20.65\\
19.& NGC 3147 & SA(rs)bc & 0.113 & 0.103 & 0.82 & 40.9 & -21.73\\
20.& NGC 6384 & SAB(r)bc & 0.112 & 0.072 & 0.72 & 26.6 & -21.31\\
21.& NGC 4030 & SA(s)bc & 0.122 & 0.105 & --- & 25.9 & -20.27\\
22.& NGC 3166 & SAB(rs)0/a & 0.126 & 0.082 & 0.93 & 22 & -20.22\\

\hline
\end{tabular}
\end{center}
\end{table}

\begin{table} 
\begin{center}
Table 1: Sample:Continued\\
\vspace{1ex}
\begin{tabular} {rlllllll}
\hline
& Galaxy & Type & S(J) & S(R) & (B-V) & D (Mpc) & M$_B$ \\
\hline
\hline
23.& NGC 3953 & SB(r)bc & 0.126 & 0.08 & 0.77 & 17.0 & -20.63 \\
24.& NGC 5985 & SAB(r)b & 0.128 & 0.103 & 0.77 & 39.2 & -21.42\\
25.& NGC 3486 & SAB(r)c & 0.14 & 0.11 & 0.52 & 7.4 & -18.23\\
26.& NGC 4123 & SB(r)c & 0.14 & 0.09 & 0.6 & 25.3 & -20.25\\
27.& NGC 3368 & SAB(rs)ab & 0.16 & 0.11 & 0.86 & 8.1 & -19.62\\
28.& NGC 5371 & SAB(rs)bc & 0.163 & 0.11 & 0.7 & 37.8 & -21.57\\
29.& NGC 3631 & SA(s)c & 0.163 & 0.12 & 0.58 & 21.6 & -20.69\\
30.& NGC 3596 & SAB(rs)c & 0.163 & 0.10 & --- & 23.0 & -20.31\\
31.& NGC 3344 & SAB(r)bc & 0.163 & 0.118 & 0.59 & 6.1 & -18.47\\
32.& NGC 6118 & SA(s)cd & 0.17 & 0.13 & 0.76 & 25.4 & -20.62\\
33.& NGC 3893 & SAB(rs)c & 0.177 & 0.13 & --- & 17.0 & -20.27\\
34.& NGC 4136 & SAB(r)c & 0.177 & 0.135 & --- & 9.7 & -18.23\\
35.& NGC 3810 & SA(rs)c & 0.177 & 0.155 & 0.58 & 16.9 & -20.11\\
36.& NGC 5364 & SA(rs)bc & 0.18 & 0.145 & 0.64 & 25.5 & -21.17\\
37.& NGC 5248 & SAB(rs)cd & 0.18 & 0.14 & 0.65 & 22.7 & -21.07\\
38.& NGC 4487 & SAB(rs)cd & 0.185 & 0.135 &  --- & 19.9 & -20.19\\
39.& NGC 3938 & SA(s)c & 0.191 & 0.149 & 0.52 & 17.0 & -20.26\\
40.& NGC 6015 & SA(s)cd & 0.20 & 0.165 & 0.57 & 17.5 & -20.05\\
41.& NGC 5585 & SAB(r)c & 0.231 & 0.162 & 0.49 & 7.0 & -17.96\\
42.& NGC 3726 & SAB(rs)cd & 0.234 & 0.188 & 0.58 & 17.0 & -20.35\\
43.& NGC 2715 & SAB(rs)c & 0.234 & 0.163 & 0.56 & 20.4 & 20.08\\
\hline
\end{tabular}
\end{center}
\end{table}
\end{document}